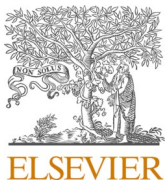
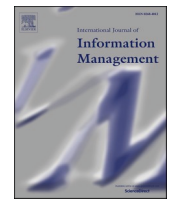

Research article

# The dark side of the metaverse: The role of gamification in event virtualization

Carlos Flavián *, Sergio Ibáñez-Sánchez, Carlos Orús, Sergio Barta

*Department of Marketing Management and Market Research, University of Zaragoza, Zaragoza, Spain*

ABSTRACT

The virtualization of cultural events in the metaverse creates opportunities to generate valuable and innovative experiences that replicate and extend in-person events; but the process faces associated challenges. In the absence of relevant empirical studies, the aim of this article is to analyze the positive and negative aspects of the user experience in a cultural event held in the metaverse. A mixed-methods approach is employed to test the proposed hypotheses. The results from three focus groups demonstrated the difficulty that users face in focusing their attention on the main elements of the metaverse, and the inability of this virtual sphere to convey the authenticity of a cultural event. Based on these findings, a metaverse-focused quantitative study was conducted to examine whether perceived gamification mitigate the negative effects of users failing to pay attention in their metaverse experiences. When users increased their attention levels, their ability to imagine the real experience and their perceptions of the authenticity of the cultural event increased, which produced positive behavioral intentions. This is one of the first studies to empirically analyze the tourist experience in the metaverse; managers and policymakers can benefit from the results to hold valuable virtual cultural events.

## 1. Introduction

The Covid-19 pandemic exposed the vulnerability of many sectors to unexpected events (He et al., 2021). This underlines the importance, in terms of creating and maintaining resilient industries, of developing effective virtualization of services with a strong experiential nature, such as tourism (Schiopu et al., 2021). In this context, the events industry (including cultural, business and entertainment sub-typologies; Getz & Page, 2016) is expected to grow at an annual rate of 13.7% and reach a value of nearly $3000 billion in 2031 (Researchdive, 2023). The virtualization of events is paramount for creating innovative and valuable experiences and protecting the sector from adverse circumstances (Yung et al., 2022).

Virtualizing events can help organizations reach wider audiences and engage with communities in novel ways; it can allow individuals to explore new cultures from the comfort of their homes, transcending physical boundaries (Yung et al., 2022). Previous research has analyzed how immersive technologies/virtual worlds (e.g., Yung et al., 2022) can be applied in the events industry. The metaverse is a technological revolution that provides highly immersive, social and interactive experiences that can, in a virtual setting, replicate and extend many aspects of in-person events (Dwivedi et al., 2022; Wong et al., 2023).

There are successful examples of virtual events being held in the metaverse (The Verge, 2020), yet others have attracted low audiences (Business Insider, 2022). Thus, while this new technology provides opportunities, there are challenges that must be faced in creating valuable experiences in the metaverse. Among these are the great number of new virtual environments that hinder the development of interoperability among these virtual spaces (Richter & Richter, 2023), and legal and privacy issues (Dwivedi et al., 2022); these may result in a restriction of the real-time multisensory social interactions that deliver satisfactory and engaging experiences (Hennig-Thurau et al., 2023). In the specific context of cultural events, conveying intangible aspects, such as values, traditions and spirituality, can be challenging.

Due to the novelty of the metaverse as a research concept, academic literature on its application to the customer experience is scarce (Dwivedi et al., 2023a). Existing research is mostly conceptual in nature (e.g., Buhalis et al., 2022, 2023; Dwivedi et al., 2022; Go & Kang, 2023; Gursoy et al., 2022; Mladenović et al., 2023; Richter & Richter, 2023; Wong et al., 2023). Among the few empirical exceptions, in the tourism industry Tsai (2022) found that users' perceptions of holistic presence (spatial, social and self-presence) in the metaverse affected (directly and

---

indirectly through holistic happiness) their behavioral intentions to visit a real destination. Thus, empirical studies, that take into account both positive and negative aspects, need to be undertaken to assess the true potential of the metaverse in terms of the customer experience (e.g., Buhalis et al., 2023; Dwivedi et al., 2023b; Hennig-Thurau et al., 2023; Wong et al., 2023).

Responding to the research gaps in the literature, the present study is one of the first empirical analyses of the user experience in a cultural event held in the metaverse. This research aims to: identify the positive and negative aspects of the user experience in a real cultural event held in the metaverse; analyze how the barriers to implementation can be overcome; examine the impact of the metaverse experience on the viewer's ease of imagining the event and the transmission of authenticity, which may subsequently affect behavioral intentions toward the event. The study draws on the theories of selective attention (Treisman, 1964) and affect as information (Schwarz, 2012), and adopts a mixed-methods approach, to qualitatively explore and quantitatively confirm the proposed hypotheses. The findings contribute to the understanding of the user experience in the metaverse and are the basis for specific proposed actions that may improve the design of virtual cultural events.

## 2. Theoretical development

### 2.1. The virtualization of tourism events

Events have been defined as one-time occasions, outside the regular programs or planned activities of a sponsoring entity, that offer users opportunities to live a leisure, social or cultural experience beyond their everyday lives (Getz & Page, 2016). Different types of tourism events exist, including business-focused (e.g., MICE, government and market fairs), entertainment (e.g., concerts, award ceremonies), sports (e.g., professional leagues/tournaments) and the cultural (e.g., festivals, religious rites) (Getz & Page, 2016). The COVID-19 pandemic prompted organizers to virtualize events, turning the virtual mode into almost second nature for these entities (Dwivedi et al., 2022; Yung et al., 2022). 'Virtualization' is the use of digital platforms (e.g., virtual reality) to create immersive and interactive experiences.

Event virtualization has benefits for organizers and participants. Virtualization provides organizers with opportunities to attract great numbers of participants to their events, yet avoids the virtual events cannibalizing the physical event experience; rather, the virtualization can complement and extend the physical event (Pearlman & Gates, 2010). Organization costs can be reduced (e.g., rental fees, travel costs of the speakers, catering), and these freed-up resources can be used to improve other aspects of events. Finally, organizers can track and collect data about attendance, participation and engagement, which is useful for quickly responding to participants' queries, to enhance the value of events for the attendees and to improve future events (Pearlman & Gates, 2010).

Virtualization allows many individuals to "attend" events that they could not attend otherwise (due to geographical, budgetary and/or time restraints). Participants save money on travel, accommodation and living expenses, and they can use their time more efficiently by avoiding trips (Kshetri & Dwivedi, 2023; Pearlman & Gates, 2010; Wong et al., 2023). Virtual events are more flexible than physical events because participants can attend sessions both synchronously and asynchronously (e.g., by viewing recordings). In alignment with the United Nations' Sustainable Development Goals, virtual events favor the inclusion of people with disabilities (Go & Kang, 2023). Finally, engagement is increased as the associated platforms usually provide highly interactive tools (e.g., chat rooms, polls; Yung et al., 2022).

Tourist destinations are increasingly using virtual events to engage with potential clients and showcase their unique cultural and natural attractions (Yung et al., 2022). By enhancing access, engagement and sustainability, virtual events can contribute to a more inclusive, innovative and responsible tourism model. The metaverse provides a new and exciting avenue for event organizers to engage with audiences (Dwivedi et al., 2022). Yung et al. (2022) argued that the metaverse can create powerful experiences by providing high social presence (i.e., a subjective sense of being together; Cummings & Wertz, 2023), which has important implications for the tourism events sector.

### 2.2. The metaverse: opportunities and challenges for virtual cultural events

The metaverse is, today, considered one of the technologies with the greatest potential (Buhalis et al., 2023; Hennig-Thurau et al., 2023; Ioannidis & Kontis, 2023). The metaverse can be defined as a persisting and continuous multi-user realm that combines physical reality with digital virtuality, using advanced technologies, such as Virtual Reality (VR) and Augmented Reality (AR), to create multisensory interactions between individuals, digital objects and virtual environments. The metaverse has generally been conceptualized as a completely generated virtual environment (Mladenović et al., 2023) or as an extended physical reality enriched by immersive technologies (Rauschnabel et al., 2022): the present study adopts the definition of the metaverse as being a connected network of immersive environments on multi-user platforms that facilitate real-time embodied communication and dynamic engagement with digital artifacts (Mystakidis, 2022).

Immersive technologies appear to be the main gateway to the virtual environment of the metaverse (Buhalis et al., 2022; Kim et al., 2023; Richter & Richter, 2023). Immersive technologies include VR (i.e., computer-generated 3D environments in which users are immersed and where they can navigate and interact), AR (i.e., digital elements superimposed over the actual user view) and pure mixed reality (PMR) (i.e., blended digital and physical elements that allow users to interact with both in real-time) (Flavián et al., 2019). The social aspect is another key component of the metaverse: users are represented by avatars which interact with other avatars in a shared, synchronous virtual environment (Ball, 2022; Han et al., 2023). In addition, the metaverse connects several virtual worlds and experiences that persist over time, even when users are not connected. Persistence and interoperability are two important aspects of the metaverse (Richter & Richter, 2023); users should be able to move elements (e.g., avatars, virtual assets) seamlessly from one platform to another (Ball, 2022; Buhalis et al., 2023). In addition to other relevant characteristics of the metaverse (e.g., decentralization, i.e., it operates independent of any central, overarching authority; Mancuso et al., 2023), and the legal and privacy issues that must be addressed by all stakeholders to make the vision of the metaverse a reality (Dwivedi et al., 2022), the current state of the key elements of technology, social interaction, persistence and interoperability, pose challenges for the development of the metaverse (Richter & Richter, 2023).

The metaverse presents unique opportunities for the virtualization of cultural events. As previously noted, users can connect to the experience from anywhere in the world (Gursoy et al., 2022), and platforms provide diverse cultural experiences (Pearlman & Gates, 2010). The metaverse can also provide effective pre-experiences that encourage users to visit cultural events in the future (Buhalis et al., 2022; Dwivedi et al., 2022) and make decisions with more confidence (Orús et al., 2021). Furthermore, the metaverse provides alternative ways to learn about cultural events (Go & Kang, 2023) and offers a personalized experience that engages users (Buhalis et al., 2023).

However, cultural events in the metaverse face serious challenges. As the metaverse evolves, so does the number of new virtual environments, which hinders its persistence and the interoperability that exists among virtual spaces, key characteristics of the metaverse (Richter & Richter, 2023). Furthermore, event organizers can gather and analyze large amounts of user data, so there is a risk that users' personal information may be exploited and/or leaked to third parties (Dwivedi et al., 2022). Users may also engage in malicious and immoral behaviors (e.g., insulting, harassing), which would negatively affect the tourist experience (Dwivedi et al., 2022). These factors might also hinder the real-time multisensory social interactions that deliver satisfactory and engaging experiences (Hennig-Thurau et al., 2023).

The transmission of a tourism event's authenticity is a critical issue. Authenticity is a universal value that motivates users to visit distant places (Cohen, 1988). Authenticity is a complex concept, the meaning of which has been much debated over the past five decades in the tourism literature (de Andrade-Matos et al., 2022; Reisinger & Steiner, 2006). Epistemologically, the present study concurs with the postpositivist stance that maintains that "objectivity, although desirable, can only be approximated" (Kolar & Zabkar, 2010, p. 654). Thus, authenticity is related to the individual's perceptions and experiences of an object or event, rather than being an objectivistic determination formed by independent judges or experts (Reisinger & Steiner, 2006). Wang (1999) argued that tourists can perceive objects (or events) as authentic even though the objects are, in fact, totally inauthentic. This may be especially important in the metaverse given that virtual events try to replicate and extend genuine, physical events; the object (i.e., the cultural event) is not authentic, yet users may still perceive it to be authentic. In the present study, authenticity is accepted as being the "enjoyment and perceptions of how genuine the experiences are" (Kolar & Zabkar, 2010, p. 655).

In a constructivist approach, researchers take the view that perceived authenticity has two dimensions, object-based and existential authenticity (e.g., Atzeni et al., 2022; Kolar & Zabkar, 2010). Object-based authenticity relates to how users perceive the genuineness or accuracy of the representations of observable elements (e.g., artifacts, architectural styles) (Atzeni et al., 2022). Existential authenticity is the perception that an activity-related experience is authentic (e.g., atmosphere, historical background, connection to culture) (Kolar & Zabkar, 2010).

Advanced techniques (e.g., 3D rendering, photogrammetry) may make it possible to transmit object-based authenticity in the metaverse; however, the transmission of existential authenticity may be more challenging. The present study aims to empirically address this issue. Specifically, we explore users' motivations and experiences in cultural events in the metaverse to identify positive and negative aspects. Based on an exploratory analysis, an online study was carried out to establish whether users' inability to pay focused attention in the metaverse can be mitigated by using gamification elements, which would aid in the transmission of existential authenticity and enhance visit intentions.

## 3. Overview of the studies

This research uses a two-stage, mixed-methods approach. In the first, exploratory stage, three focus groups were established. Focus groups are appropriate for defining key variables and determining their interrelationships in the early stages of research (Kidd & Parshall, 2000) (which applies to the study of virtual tourism events in the metaverse). In the second, confirmatory stage, a quantitative study was carried out in which participants experienced a cultural event in the metaverse and, thereafter, answered a related questionnaire.

There are several reasons for using mixed-methods approaches in research (Bryman, 2006). First, using qualitative and quantitative analyses to triangulate and corroborate findings provides additional validity. Second, mixed methods enable researchers to obtain a comprehensive overview of a research issue that allows them to develop an initial framework. Third, qualitative data can be used to generate hypotheses, which can then be tested by quantitative research, in accordance with the confirm and discover rationale. Finally, on the basis of the completeness rationale and the utility rationale, combining two approaches provides an applied focus that can help researchers make practical proposals.

Maier et al. (2023) cross-sectional data collection guidelines were followed. First, a sampling strategy was adopted to collect a representative sample to fit the study's context. Second, sample size requirements were determined to ensure the suitability of the model proposed in the quantitative study. To establish a medium effect size, with a power level of 0.80 and a significance level of 0.05, the required sample size is 150 (Soper, 2023). The sample size in our study, 219, thus exceeded the minimum requirements. Third, the cross-sectional data was introduced in a mixed-methods design.

## 4. Study 1: qualitative research

### 4.1. Methodology

Three focus groups were conducted (between 5 and 8 participants per session) to examine specific issues (Krueger, 2014). The sessions lasted between 60 and 90 min. The participants were recruited in Spain, following a non-probabilistic, purposive approach. The composition of a focus group should have a certain degree of homogeneity to avoid huge differences in opinion emerging, but it should also be diverse enough to promote discussion and generate useful information (Phillippi & Lauderdale, 2018). As prior knowledge of cultural events can influence participants' perceptions and evaluation of an experience (Lobuono et al., 2016), we selected people with similar levels of knowledge about the cultural event under consideration, but with different characteristics in terms of age, gender, and willingness to adopt new technologies. The focus groups were run until the saturation criterion was met (Malterud et al., 2016). The composition of the focus groups is shown in Table 1.

The focus groups were designed to obtain the most valid and meaningful data (Kidd & Parchsall, 2000). A few days before the focus groups took place, their participants were asked to access a particular virtual space (a well-known, traditional Spanish cultural event) to familiarize themselves with the navigation (see Fig. 1).[1] This cultural event, held annually in the city of Zaragoza, attracts hundreds of thousands of visitors worldwide. The goal of the study was explained to the participants before the sessions (see Appendix A for an outline). The sessions began with a presentation of the topic to the participants, and

**Table 1**
Focus group participants.

| Part | Focus Group | Age | Profession | Gend |
| --- | --- | --- | --- | --- |
| 1 | 1 | 23 | Business student | F |
| 2 | 1 | 25 | Physics student | M |
| 3 | 1 | 23 | Medical student | F |
| 4 | 1 | 23 | Teaching student | F |
| 5 | 1 | 39 | Computer scientist | M |
| 6 | 1 | 24 | Estate agent | F |
| 7 | 1 | 21 | Finance student | F |
| 8 | 1 | 26 | Consultant | M |
| 9 | 2 | 59 | Houseperson | F |
| 10 | 2 | 18 | Veterinary student | F |
|  | 2 |  |  |  |
| 11 | 2 | 24 | Business student | F |
| 12 | 2 | 23 | Medical student | F |
| 13 | 2 | 18 | Engineer student | M |
| 14 | 2 | 61 | Civil servant | M |
| 15 | 2 | 23 | Business student | F |
| 16 | 3 | 29 | PhD Student | F |
| 17 | 3 | 28 | PhD Student | F |
| 18 | 3 | 22 | Law Student | M |
| 19 | 3 | 28 | PhD Student | M |
| 20 | 3 | 32 | Assistant Professor | M |

Notes: Part=Participant; Gend=Gender; F=Female; M=Male.

---

[1] The event virtualization was carried out by the Imascono company. This virtual recreation was recently acknowledged by the European Commission as one of the most advanced instances of cultural reenactment in the virtual sphere (Hupont Torres et al., 2023).

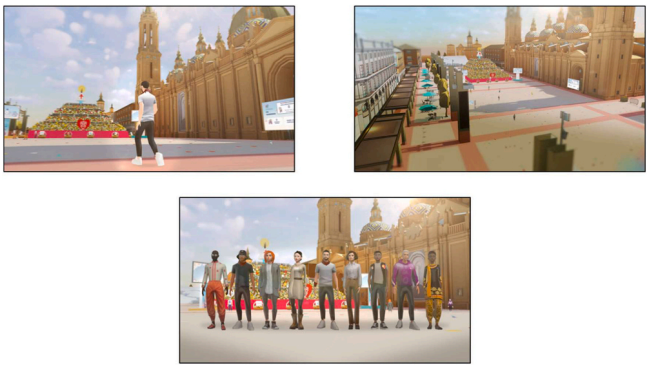

**Fig. 1.** Cultural event held in the metaverse employed in the study. *Source: Imascono Art S.L. (*https://imascono.com/*).*

then several questions were posed to obtain the perspectives of the groups on the topic. During the focus groups, the participants again accessed the cultural event, but on this occasion they were explicitly requested to play some games that had been introduced into the virtual environment (Appendix A). The focus group discussions were recorded and transcribed.

*4.2. Results*

Three independent researchers, specializing in immersive technologies, coded the transcriptions of the focus group discussions to identify core themes and illustrative statements. The goal was to determine the main themes of the discussions by identifying commonalities and by grouping key aspects of each theme. After coding the transcription into predetermined categories, the coders discussed discrepancies between their analyses until they reached a consensus.

Table 2 summarizes the main themes discussed in the focus groups and provides illustrative statements. The first questions posed were about participants' perceptions of the virtualization of events in the metaverse. Overall, event virtualization was not perceived as suitable for events with a high recreational component. To minimize the loss of value derived from such events, using devices with multi-sensory experiences may be appropriate. In addition, using graphic effects or other elements unavailable in the physical experience might enrich virtual events. Finally, as well as taking into account the nature of the event (e. g., recreational vs work-related), the characteristics of the target audience should be considered when deciding whether to virtualize a cultural event. It may be that virtualization is more appropriate for audiences with experience of immersive technologies and for innovative users.

The second group of questions evaluated the participants' previous experience of the metaverse. There was a strong consensus that visitors to the metaverse need to be given clear operating instructions. The participants often "felt lost", unsure of what to do and did not know whether they were browsing well or if they were missing some aspects of the experience due to their lack of knowledge. This negative side of the experience provoked negative feelings, such as confusion, anxiety and tiredness. However, these aspects can be mitigated with the implementation of gamification elements, which help participants and provide entertaining experiences by providing challenging activities.

The last set of questions addressed the potential of the metaverse to convey the authenticity of the cultural event. The transmission of the authenticity of the cultural event was highlighted as an important challenge of the metaverse. Apart from visual aspects, such as the adaptation of the avatars to the characteristics of the event, it is essential to include other multi-sensory aspects (e.g., typical smells, music and tactile sensations of the event) to convey an event's existential authenticity. For example, displaying the most characteristic heritage elements in the virtual environment, and the customization of avatars in terms of

**Table 2**
Results of the focus groups.

| Themes | Description | Example participant statements |
|---|---|---|
| **Event virtualization** | Interesting for virtualizing work-related events, such as conferences. This avoids having to gather attendees in the same city, travel time, accommodation costs. | *"There are events, such as conferences, that can be of interest to people who are used to working this way. Instead of all of the attendees being in the same city, it can be interesting to attend remotely. However, for many social events, such as concerts, it's not the same. You need to feel the people next to you jumping; and vibrations from speakers are not replaceable. Perhaps, in the future, with the use of devices, such as haptic gloves or vests, this could be solved."* (P5, 39 years old, male). |
| | Recreational events (such as concerts) lose value if they are virtualized. Devices (e.g., virtual reality headsets) that provide multi-sensory experiences would improve the virtual experience. | *"During quarantine, I "attended" a Fortnite concert. At that time, I was playing the video game and it seemed like a good option due to the global situation. Of course, a live concert is better, but in situations like this, it allowed me to enjoy an event and a series of graphic effects that you could not see in person."* (P2, 25 years old, male). *"I think you can visit a museum in a virtual tour and you would be less concerned about the people there. In the case of a concert, I think the environment where it takes place and the people and the friends you go with are more important. Therefore, I think it is more complicated to virtualize this type of event."* (P9, 59 years old, female). |
| | The target audience of the event should be taken into account. If it is an audience used to employing new technologies, problems related to the technology would be reduced and the virtualization experience would be improved. | *"I think it would help the event's virtualization if it was targeted at people who are more accustomed to new technologies."* (P19, 28 years old, male). |
| **Metaverse experience** | Metaverse users are not quite sure what to do when they access the environment. To improve their experience in the metaverse, it would be advisable to provide users with instructions about the activities they can undertake or a mini tutorial the first time they access the space. | *"I went into the metaverse wanting to know more about it, but I did not know what to do. There was no one and no information to guide me or tell me what to do. I did not really understand what my objective was there."* (P3, 23 years old, female). *"I was not very sure what the goal was. You do not know how to browse, you do not have much interest and you do not spend much time in it. I did not know what I could do to have a good experience."* (P8, 26 years old, male). *"As the instructions given at the beginning are limited, I was anxious because I didn't know if I was doing it right. I think that having a little knowledge of how it works, and having a* |

**Table 2** (*continued*)

| Themes | Description | Example participant statements |
|---|---|---|
| | | *bit of experience with the platform, improves the experience, as you know how to move faster and understand the activities you can do.*" (P14, 61 years old, male). |
| | The experience can be engaging and interesting for short periods of time. Once users have explored the entire environment and completed different activities, it no longer provides entertainment, and they would not access to the metaverse again. | "*The current metaverse technology provides entertainment for a while. However, finally, the virtual space itself is limited, and once you have moved around the environment a bit, it is not so much fun anymore. I got tired of navigating in the environment afterward, not having any activities to do.*" (P18, 22 years old, male). |
| **Gamification in the metaverse** | Gamification in the metaverse improves the experience, helping users learn what to do in the virtual environment. | "*The games make the metaverse more entertaining, they help you to better understand the simulation of the event.*" (P10, 18 years old, female). |
| | Quizzes help users learn about the cultural event, making the experience more useful. | "*The quiz entertained me. It is like a competition with yourself, it allowed me to learn new things. Being there and taking advantage of the opportunity and including this kind of aspect adds value to the experience.*" (P13, 18 years old, male). |
| | Challenging games improve users' web navigation, making them feel that they know what activities to undertake in the environment. These games also provide entertainment, thus enhancing the experience. | "*Showing the cultural event with these challenges makes you have fun, learn about the cultural event and interact with the whole environment. This enhances the experience even more.*" (P17, 28 years old, female). |
| **Authenticity** | Conveying authenticity through the metaverse is challenging. | "*In my case, I do not have a very strong feeling for this cultural event. I do not feel that it is a recreation of the event, but it is more like a game that takes place in an environment that you recognize.*" (P2, 25 years old, male). |
| | Displaying the most characteristic heritage elements in the virtual environment is key to achieving a sense of authenticity. | "*For me, the more traditional features of the event were missing. The flowers, the smell of the flowers, the typical clothes, the crowds of people, the key aspects that define what the event is.*" (P9, 59 years old, female). |
| | The customization of avatars, clothing and other distinctive aspects of the event is also important to convey authenticity. | "*The perception of authenticity would be better if, instead of seeing animated representations of buildings and the fountain, would be to view a representation of the real environment in which I move through with my avatar […]. In addition, it is key that the characteristic heritage elements also appear in the virtual world.*" (P16, 29 years old, female). |
| | Visual aspects must be complemented with those perceived through other senses. The sound / music characteristic of the place and | "*The experience would be much more authentic if I could dress my avatar in the traditional clothes worn on the day and if it could do the* |

**Table 2** (*continued*)

| Themes | Description | Example participant statements |
|---|---|---|
| | the smell of the cultural event, among others. | *traditional dance.*" (P17, 28 years old, female). |
| | Gamification can enhance the perceived authenticity of the metaverse because it guides users to perform certain activities, improving their affect during the experience. | "*I found the experience more entertaining because I no longer felt like I was just walking around in a virtual world without knowing what to do. For me, gamification made the experience more fun and allowed me to find out more things related to the cultural event. I found this experience more interesting.*" (P18, 22 years old, male). |

clothing, and of other distinctive aspects of the event, are key to transmitting authenticity.

The results of this exploratory study provide a user perspective on the virtualization of cultural events in the metaverse. Although positive aspects were identified, several issues must be taken into account, such as the nature of the event (e.g., work-related versus recreational) and the target audience. Two important concerns were raised. First, the users often felt lost: the virtual environment contains many elements and activities, thus, without proper guidance, the users could not focus on 'living the experience', which produced negative experiences and hindered the transmission of the authenticity of the event. Second, gamification appeared to provide purpose and improved the experience by diminishing its negative consequences. The following study seeks to quantitatively confirm these findings.

## 5. Study 2: quantitative study

### 5.1. Theoretical basis

The results of the qualitative study showed that users can be overwhelmed in the metaverse because of its vast amount of information, visual objects and activities to undertake. This prevents them from paying attention to important information about events. Selective attention theory proposes that the ability to focus on specific information and filter out irrelevant information is crucial to cognitive processing (Treisman, 1964). Treisman (1964) argued that the human brain can process only a limited amount of the sensory information it is presented with at any given time. Selective attention allows individuals to direct their cognitive resources to the most important information, while ignoring the rest. Consequently, sensory inputs which are not selected are paid less attention and are processed with less intensity.

Previous research into online browsing found that selective attention theory is a suitable framework for understanding the cognitive mechanisms underlying attention and perception in online environments (Wickens, 2021). If users perceive online browsing as difficult, complex and/or unfamiliar, they will assign it with more mental resources and, in consequence, will have little attention left to devote to anything irrelevant to the task in hand, such as looking at the web environment (Flavián et al., 2010; Guo et al., 2020; Liu et al., 2018).

If users cannot focus their attention on, and navigate within, virtual environments in a simple and intuitive way, they are likely to undergo unpleasant experiences. Affect as information theory proposes that individuals' emotions are strong determinants of their cognitive evaluations, and that they automatically integrate these states into their decision-making processes (Schwarz, 2012). Consumers evaluate objects (e.g., products, experiences) on the basis of how they feel about them. This theory has been successfully applied to user experiences in virtual and immersive technologies (e.g., Whittaker et al., 2021; Zanger et al., 2022). In sum, the theory of selective attention and affect as information theory serve as the theoretical basis of the present study. If, because they cannot focus their attention, users develop negative states during their experiences in the metaverse, these negative feelings can influence their cognitive evaluations of the experience.

*5.2. Hypotheses development*

The conceptual model used in Study 2 is based on the results of the qualitative analysis, and the research hypotheses are based on the theory of selective attention, affect as information theory and evidence taken from previous literature. Specifically, Study 1 found that users felt lost when navigating in metaverse experiences, particularly when they were not accustomed to the environment, which resulted in them failing to pay focused attention on living and enjoying the experience. A lack of focused attention generates negative emotions that affect the overall evaluation of the experience and the perceived authenticity of the event. However, the focus group participants reported that gamification improved their experiences of the virtual event. Gamification elements helped the participants to navigate and focus attention, which reduced the incidence of negative emotions and led to a more positive evaluation of the experience.

Consequently, it is proposed on the basis of the results of the research model that the lack of attention paid by users to important information about cultural events generates a negative effect in them that influences their ability to imagine the event and accept its authenticity, which generates negative behavioral intentions toward the event. It is also proposed that the detrimental effects of lack of attention can be mitigated by the inclusion of game elements in the metaverse.

Selective attention theory proposes that lack of attention can impact on affective processes (Treisman, 1964). When users are unable to pay attention to relevant information in their environments, they may experience increased stress and anxiety about their actions and may become frustrated with their inability to find what they are looking for (Zheng & Ling, 2021). Consequently, they can suffer negative affect, which is subjective distress and unpleasant engagement that is associated with a variety of adverse mood states, including frustration, discomfort and feeling lost (Qi et al., 2015). The metaverse is a vast and complex source of information which can overwhelm its users (Mystakidis, 2022). As its users are not accustomed to processing such a high volume of information, they may suffer from cognitive overload, which can cause negative affect (Zhang et al., 2016). In addition, as the metaverse emerged only recently, its users may not pay attention to its more important elements because they do not know what to do, and how to interact, within it. These aspects can lead users to feel lost, which can further impair attention levels and lead to negative affect:

**H1**.   : Lack of focused attention has a positive effect on negative affect.

Affect as information theory proposes that affective states generated during a metaverse experience in a virtual cultural event may influence how users perceive and evaluate the experience. Specifically, negative affect can have a significant impact on imagery fluency, which is a subjective experience related to the individual's capacity to imagine what an object looks like in reality (Orús et al., 2017; Ruusunen et al., 2023). Previous research into immersive technologies has shown that affective states influence subjective experiences (Kowalczuk et al., 2021). Emotions have an important effect on individuals' imaginations. When individuals experience negative affect, their attentional resources may be directed towards the source of their distress, making it more difficult for them to focus on other tasks, including using their imaginations (Hong et al., 2004). The frustration/annoyance felt by users during their interactions with technologies creates cognitive interference that draws on their cognitive resources (Park et al., 2021). As a consequence, their ability to process information and engage in imaginative thinking is hindered. When users experience negative feelings while in the metaverse, their ability to imagine cultural events is diminished:

**H2a**.   Negative affect has a negative effect on users' ease of imagination of the cultural event.

Negative affect can also impact on the metaverse's capacity to transmit an event's authenticity. Existential authenticity is the perception that an activity-related experience is authentic in terms of its atmosphere, historical background and connection to the relevant culture (Kolar & Zabkar, 2010). Emotional responses to products or services influence users' perceptions (Rathore & Ilavarasan, 2020). Negative emotions are evoked when individuals feel distressed at being in an unpleasant situation (Qi et al., 2015). Negative states can affect users' perceptions while they are in the metaverse, which can reduce their ability to perceive its authenticity (Illouz, 2017). In particular, negative affect can lead individuals to lose touch with the core values and main aspects (e.g., history, atmosphere) of an event, which disconnects them from the cultural experience. Consequently, negative emotions can diminish users' perceptions of the existential authenticity of an event:

**H2b**.   Negative affect has a negative effect on users' perceptions of the existential authenticity of the cultural event.

Negative affect can have a significant impact on behavioral intentions, defined in this research as the individual's plans or motivations to engage with a cultural event (Ratnasari et al., 2021). Negative affect can lead individuals to adopt avoidance behaviors, that is, they avoid situations they perceive as being unpleasant (Stasiewicz & Maisto, 1993). When individuals are in a place and experience negative affect, they are less willing to develop positive behaviors toward that place (Schwarz, 2012). This can be extrapolated from virtual to real environments: the emotions evoked in a virtual experience can persist beyond the virtual setting and lead users to change their behaviors in offline settings (McLeod et al., 2014). Thus, if users experience negative feelings when they attend cultural events in the metaverse, this negative state can transfer to, and influence, real-world behaviors. Formally, it is proposed that negative affective states will decrease the individual's intention to develop positive behaviors toward the cultural event:

**H3**.   : Negative affect has a negative effect on behavioral intentions toward the cultural event.

The ease of with which they imagine what an event would be like can affect users' perceptions and behaviors (Orús et al., 2021). When individuals easily imagine themselves experiencing an event, their sense of the existential authenticity of the event increases, even if they are not physically present (Atzeni et al., 2022). Existential authenticity is perceived when individuals cognitively align their mental self-representations with being present and observing a cultural event through the metaverse. When participants immerse themselves in the virtual experience of the cultural event, they can form a vivid mental image of how the real-life experience would unfold (Orús et al., 2021). Going through this imaginative process makes the virtual experience tangible, which makes individuals more able to appreciate the event's cultural elements, including its history, customs, heritage and atmosphere (Atzeni et al., 2022). Thus, if individuals attending a cultural event in the metaverse can easily imagine the place and the event, they will be more likely to accept its existential authenticity:

**H4a**.   Users' ease of imagination has a positive effect on the existential authenticity of the cultural event.

Ease of imagination can also create sensory experiences, such as visualizing oneself performing a particular behavior and/or imagining the positive outcomes that may result from engaging in that behavior (Orús et al., 2021). In the context of cultural events, as ease of imagination increases users' interest, individuals may develop a desire to convert the virtual experience into a physical visit. Immersive technologies allow users to better imagine how a tourism experience would be, which may drive them to undertake positive behaviors in real life (Flavián et al., 2021). Similarly, in the metaverse, potential travelers can use their imaginations to envision tourism offerings and the value/benefits they might derive from visiting a destination, which may shape their subsequent behaviors (Dwivedi et al., 2022). Thus, if users can imagine how things would turn out at a real event, they will be more likely to develop positive behavioral intentions:

**H4b.** Users' ease of imagination has a positive effect on their behavioral intentions toward the cultural event.

Existential authenticity can play a crucial role in intentions to visit an event (Park et al., 2019). When cultural tourism experiences have high levels of existential authenticity, users experience a sense of enjoyment and well-being (Atzeni et al., 2022). Furthermore, when the values and traditions of a cultural event are successfully transmitted, users' curiosity to attend the event may be aroused (Kim et al., 2019). Previous research has argued that existential authenticity is an important dimension in explaining post-experience tourist satisfaction and loyalty (Park et al., 2019). The use of innovative technologies can support this process. Atzeni et al. (2022) demonstrated that attending a tourist attraction through non-immersive virtual reality helped users gain a deep insight into the local history, culture and atmosphere of a tourist attraction, which prompted them to visit the destination in the future. Therefore, it is proposed that the perceived existential authenticity derived from an experience in the metaverse is an important predictor of users' behavioral intentions toward a cultural event:

**H5.** : Existential authenticity has a positive effect on users' behavioral intentions toward the cultural event.

### 5.2.1. Moderating effect of perceived gamification

Gamification is the use of game elements and mechanics in non-game contexts (Mishra & Malhotra, 2021). Gamification has the potential to affect the user's selective attention (Seaborn & Fels, 2015). First, gamification increases user engagement and motivation through rewards, feedback and challenges designed to be intrinsically motivating (Cechetti et al., 2019; Koivisto & Hamari, 2019). This increased engagement can lead individuals to devote more sustained attention to the task at hand, as they are more likely to focus on activities they find rewarding and enjoyable (Cechetti et al., 2019; Saleem et al., 2022).

Second, gamification can enhance user interaction with the metaverse by providing immediate feedback, personalized instructions and opportunities to practice and experiment. This can lead them to navigate more efficiently and effectively, as in this instance individuals will not feel lost, but will focus their attention on the relevant aspects of the environment (Saleem et al., 2022). Where individuals are not able to pay attention to relevant inputs from the metaverse environment (due to the reasons discussed in the qualitative analysis), gamification can alleviate the effects of this disconcerting state that leads to negative affect. Formally, we propose that perceived gamification has a moderating effect on the relationship between lack of focused attention and negative affect:

**H6.** . The effect of lack of focused attention on negative affect weakens as perceived gamification increases.

### 5.3. Data collection and measures

An online study was carried out to test the hypotheses. The participants were recruited through a market research agency; they received an economic reward. The virtual event used in the qualitative study was employed as the stimulus in the online study. The sample of participants was drawn from the Spanish panel of the market research agency.

The study is in two parts. In the first, the participants accessed the metaverse to experience the cultural event. Based on the results of the exploratory study, information was provided to the participants about how to navigate within the virtual space, interact with objects and virtual assistants, and all the activities they might carry out (including the gamification elements). A procedure was implemented to control the participants' access to the metaverse: first, they were asked to upload a screenshot of their avatar from the virtual space into their questionnaire; second, the system registered the time the participants spent in the metaverse. Based on the results of pre-tests ran prior to the main study, and the results of the qualitative analysis, participants who spent less than three minutes in the metaverse were excluded. Third, control questions were introduced into the follow-up questionnaire to verify the participants had paid attention while they were in the metaverse. Following this screening procedure, a total valid sample of 219 participants was obtained. The sample was relatively young ($M = 31.67$; $SD = 9.84$) and balanced in terms of gender (51% female).

After the metaverse experience, the participants were presented with a questionnaire (see Appendix B). Previously validated scales were used to measure lack of focused attention (Nelson et al., 1993), negative affect (Eppmann et al., 2018), ease of imagination (Orús et al., 2017), existential authenticity (Atzeni et al., 2022), behavioral intentions (Flavián et al., 2021) and perceived gamification (Esmaeilzadeh, 2021). Seven-point Likert scales (1 = "strongly disagree", 7 = "strongly agree") were used. In addition, personal variables related to the participants' level of knowledge of the cultural event (Smith & Park, 1992), their emotional involvement (Zaichkowsky, 1994) and their perceived place attachment (Belanche et al., 2016) were measured as control variables. Finally, sociodemographic information was collected.

### 5.4. Results

#### 5.4.1. Measurement model and assessment

Analyses of the reliability and convergent validity of the scales were conducted using SmartPLS 4.0 software. The factorial loadings of the indicators exceeded the minimum recommended level of 0.70 (except one, see Appendix B; Hair et al., 2011). The composite reliability of the constructs and the average variance extracted (AVE) values were also higher than the recommended minimum levels (Hair et al., 2011) (see Appendix B). Discriminant validity was assessed based on the criteria of Fornell and Larcker (1981) and heterotrait-monotrait ratios (Kline, 2011), with both approaches returning satisfactory values (see Table 3).

#### 5.4.2. Structural model results

To test the model's hypotheses, a bootstrapping method, using SmartPLS with 5000 subsamples, was used (Hair et al., 2011). $R^2$ values (coefficient of determination) are a measure of the predictive ability of a structural model. $R^2$ values higher than 0.25, as a rule of thumb, indicate a model has moderate explanatory power (Hair et al., 2011). The results of the tests of the hypotheses and the $R^2$ values are shown in Fig. 2.

The analysis showed that lack of focused attention had a statistically positive effect on negative affect (H1 supported). The negative affect resulting from the experience had significant and negative effects on ease of imagination and perceived existential authenticity (see Fig. 3), supporting H2a and H2b. However, lack of focused attention had no statistically significant effect on behavioral intentions (H3 rejected). Ease of imagination positively influenced perceived existential authenticity and behavioral intentions toward the cultural event, supporting H4a and H4b. Finally, perceived existential authenticity had a significant and positive effect on behavioral intentions (supporting H5).

Regarding the moderating effect proposed in H6, perceived gamification in the metaverse had a statistically significant moderating effect on the relationship between lack of focused attention and negative affect (Fig. 3). As Fig. 3 shows, the impact of lack of focused attention on negative affect decreased as users' perceptions of the gamification of the environment increased (H6 supported). In addition to the proposed direct impacts, it was observed that perceived gamification significantly moderated the effect of lack of focused attention through negative affect on ease of imagination ($\beta = 0.051$, $p < 0.01$) and the effect of lack of focused attention through negative affect on perceived existential authenticity ($\beta = 0.032$, $p < 0.01$). That is, lack of focused attention had

Table 3
Discriminant validity of the scales.

| Variables | 1 | 2 | 3 | 4 | 5 | 6 | 7 | 8 | 9 | 10 | 11 |
|---|---|---|---|---|---|---|---|---|---|---|---|
| (1) Lack of focused attention | **0.817** | 0.603 | 0.585 | 0.676 | 0.503 | 0.566 | 0.316 | 0.305 | 0.240 | 0.047 | 0.160 |
| (2) Negative affect | 0.558 | **0.883** | 0.462 | 0.575 | 0.407 | 0.407 | 0.130 | 0.109 | 0.122 | 0.046 | 0.048 |
| (3) Ease of imagination | -0.515 | -0.432 | **0.908** | 0.808 | 0.719 | 0.427 | 0.256 | 0.330 | 0.229 | 0.116 | 0.102 |
| (4) Existential authenticity | -0.594 | -0.520 | 0.728 | **0.818** | 0.736 | 0.489 | 0.298 | 0.475 | 0.329 | 0.055 | 0.149 |
| (5) Behavioral intentions | -0.439 | -0.381 | 0.669 | 0.660 | **0.928** | 0.244 | 0.315 | 0.511 | 0.336 | 0.140 | 0.159 |
| (6) Perceived gamification | -0.471 | -0.345 | 0.352 | 0.398 | 0.195 | **0.829** | 0.127 | 0.185 | 0.095 | 0.108 | 0.062 |
| (7) Cultural event knowledge | -0.268 | -0.119 | 0.232 | 0.262 | 0.282 | 0.069 | **0.941** | 0.510 | 0.405 | 0.022 | 0.251 |
| (8) Cultural event involvement | -0.269 | -0.089 | 0.304 | 0.422 | 0.465 | 0.128 | 0.444 | **0.909** | 0.490 | 0.064 | 0.379 |
| (9) Place attachment | -0.209 | -0.113 | 0.218 | 0.293 | 0.310 | 0.024 | 0.361 | 0.441 | **0.900** | 0.086 | 0.347 |
| (10) Gender (0 =Female) | -0.006 | 0.030 | 0.109 | 0.053 | 0.134 | 0.088 | 0.021 | -0.010 | -0.083 | **N.A** | 0.146 |
| (11) Age | -0.147 | 0.001 | 0.101 | 0.138 | 0.152 | 0.050 | 0.234 | 0.354 | 0.334 | -0.146 | **N.A** |

Notes: N.A = Not Applicable. The diagonal elements (in bold) are the square roots of the AVEs. Above the diagonal elements are the HTMT values. Values below the diagonal elements are the inter-construct correlations.

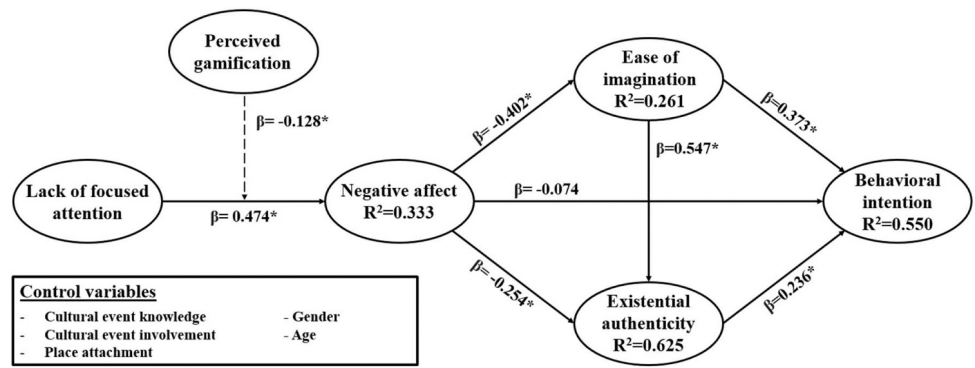

Fig. 2. Results of the study. *Note: solid lines: direct effects; dotted line: moderating effect; * : significant at 0.01 level.*

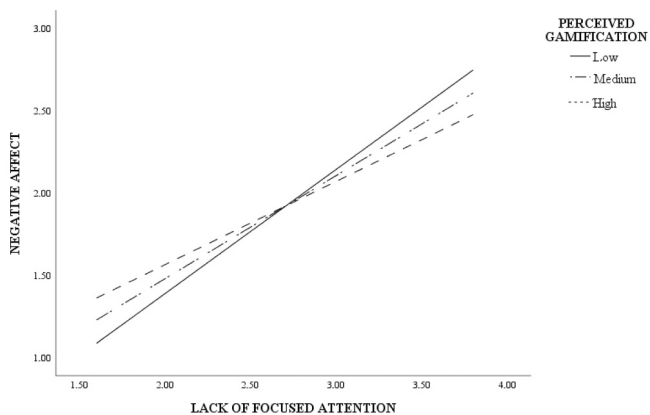

Fig. 3. Moderating effect of perceived gamification on the relationship between lack of focused attention and negative affect. *Note: low, medium and high levels of perceived gamification correspond to percentiles 16th (M = 4.33), 50th (M = 5.67), and 84th (M = 6.92) of the scale.*

a negative indirect effect on ease of imagination ($\beta = -0.195, p < 0.01$) and on existential authenticity ($\beta = -0.124, p < 0.01$). Perceived gamification mitigated these indirect effects.

Finally, the effects of personal factors on the dependent variables were controlled. Specifically, variables related to the event (knowledge of, and involvement with, the cultural event), the place (place attachment) and sociodemographic characteristics (gender and age) were included. The significance of the established relationships did not vary after introducing these variables. Thus, the results of the model are robust.

## 6. Discussion

The virtualization of metaverse-based events offers extraordinary opportunities. The scope of events can be extended by allowing people around the world to virtually attend events they otherwise could not. Virtualization solves the problems of distance, capacity and attendee mobility. However, it should be recognized that the virtualization of metaverse-based cultural events is challenging, and several negative barriers must be overcome to ensure advantage is taken of its unique possibilities (Dwivedi et al., 2022, 2023a). The metaverse experience must be further improved to ensure virtual cultural events more closely resemble real events.

One of the main advantages of the metaverse is its multiple, easily customizable virtual elements, which provide enormous opportunities for user interaction (Mystakidis, 2022). However, the results of this study suggest this aspect can be problematic. When users access the metaverse, they encounter a large amount of information and multiple possible interactions. This can make them feel overwhelmed, as they may not know what to do. Users need to spend time and receive clear operating instructions to avoid feeling lost in the virtual world, a sentiment which creates poor experiences.

Poor experiences derived from the user's inability to pay focused attention to important elements of metaverse events generate negative feelings that influence the users' capacity to imagine the event and hinder their acceptance of its authenticity; consequently, they do not develop positive intentions to attend future events. When users are in positive affective states, they can easily imagine events and perceive them as being authentic (Atzeni et al., 2022). Just as perceived authenticity has been shown to generate positive behavioral intentions in other generic tourism-related contexts, for example, toward destinations (Kim et al., 2020), this research has shown that they are generated also toward traditional cultural events. On the other hand, in the present study affective states were not found to have a direct effect on tourists' behavioral intentions, rather they had an indirect influence through the tourist having a positive mental predisposition toward the event. This finding is aligned with the experiential hierarchy model (Holbrook & Hirschman, 1982), which has extended its reach into the immersive technologies landscape (Kowalczuk et al., 2021).

In addition, extending previous results about the role of gamification in online contexts (Aparicio et al., 2019; Mishra & Malhotra, 2021), it would appear that gamification elements are appropriate for use in virtualized cultural events. Lack of focused attention can generate an unpleasant surfing experience. When faced with many alternatives in decision-making processes, individuals can enter negative states (e.g., confusion; Barta et al., 2023). The present study has shown that this phenomenon might be encountered in the metaverse. Gamification can prevent users undergoing poor experiences by providing them with entertainment and guidance. Gamification elements stop users feeling lost and, therefore, improve their experiences in terms of their ability to imagine events and accept their authenticity, which subsequently determines their visit intentions.

### 6.1. Theoretical implications

The present study provides a better understanding of the user experience in the novel field of the metaverse. As previously noted, the vast majority of prior research into the metaverse and its application to the tourism industry is theoretical (e.g., Buhalis et al., 2022; Go & Kang, 2023; Gursoy et al., 2022; Buhalis et al., 2023; Wong et al., 2023), with few exceptions (Tsai, 2022). In response to calls for empirical studies in this context (Buhalis et al., 2022; Dwivedi et al., 2022, 2023a), the present study is one of the first to analyze qualitatively and quantitatively the user experience in the metaverse in the field of tourism.

This research focuses on event tourism, where the metaverse offers new opportunities for delivering memorable experiences (Yung et al., 2022). The metaverse, renowned for its novelty, immersive qualities and social nature, is a dynamic, innovative digital platform that can host cultural events (Buhalis et al., 2023). Yung et al. (2022) argued that the advent of the metaverse has important implications for the events industry as it can fully replace, complement and extend the physical event experience. Taking into account the great potential of the metaverse, the present study complements the existing literature by analyzing the user experience of this digital realm and its role in cultivating interest among prospective attendees of cultural events.

Most studies on the application of new technologies focus solely on their positive impacts on the user's experience (Bunjak et al., 2021). However, the "dark side" of these technologies must also be explored to obtain a full picture of the user experience (Kotler et al., 2016). Based on selective attention theory and the affect as information theory, this research contributes to the literature by exploring (in a mixed-methods approach) the distracting nature of the metaverse. This study underlines a negative aspect of the metaverse. That is, the user's inability to focus his/her attention, due to the vast amount of information and multiple activities the metaverse provides, creates negative affect (e.g., feeling lost, frustration) that diminishes his/her ability to imagine cultural events and accept their existential authenticity; and, users' subjective evaluations of ease of imagination and existential authenticity influence their behaviors toward events.

However, there is light at the end of the tunnel: gamification elements seem to be tools effective in reducing the negative effects that can be generated by virtual experiences. The present study responds to calls for studies that analyze the impact of gamification on the user experience in the metaverse (Buhalis et al., 2023; Dwivedi et al., 2023a): when addressing gamification tasks, users focus their attention on the important content of the environment and enjoy an improved experience. Perceived gamification also aids the user's ability to imagine events and to accept their existential authenticity, both of which enhance his/her behavioral intentions.

### 6.2. Managerial implications

Tourism companies and institutions have real opportunities to showcase their products in the exciting realm of the metaverse. Nonetheless, several challenges must be faced to generate valuable experiences. There are practical examples of successful metaverse events, but there are also instances of poor attendance/investment ratios. The results of the qualitative research suggest that careful thought should be given to what event type (e.g., recreational or utilitarian) should be virtualized on the metaverse. When the event is work-focused (e.g., conferences, workshops) users see the value of using the metaverse; however, in recreational events (e.g., concerts, cultural acts) companies should strive to provide multisensory experiences that engage users in valuable experiences. This might be achieved by using sensory and emotionally stimulating embodied technologies (e.g., head-mounted displays, haptic devices). Offering content that cannot be accessed in the real world (e.g., high-quality, interactive graphics) can enrich the user's experience in metaverse-based recreational events. The target audience of the event should also be taken into account when designing virtual events. Participants familiar with new technologies (e.g., younger generations, innovative users) expend less effort in learning how to navigate and may be more prone to use them than others less familiar. This may be particularly important for events aimed at mass audiences, which are often promoted by public entities, as they need highly user-friendly and accessible environments.

In the context of cultural events, event managers should recognize the negative aspects of the metaverse, particularly when the event is a "one-time experience" created for a specific occasion. The results of this mixed-methods study showed that, in the context of metaverse-based events, the many elements and activities, and the poor guidance offered to users, make them feel lost, which prevent them from focusing on "living" the experience. This lack of focused attention produces negative affective states that subsequently translate into a lower capacity to imagine events, and accept their authenticity, and lower intentions to visit real events. Thus, event organizers are strongly encouraged to offer users assistance that can guide their experiences. Just as in physical events, signs ("entrance" and "exit"), maps, instructions and virtual staff would help users navigate in the metaverse.

The results of this research offer a potential solution that may mitigate the negative effects caused by users' inability to pay attention to important aspects of their experiences in the metaverse. Specifically, the use of gamification elements is highly recommended because they help users focus their attention on the virtual space and experience the atmosphere of the cultural event. For example, event designers might organize "scavenger hunts", hiding objects in the environment, perhaps close to specific points, for example, information panels. As well as offering purpose and an entertaining experience, this may prompt users to explore the environment, become familiar with how to use their avatars and to identify key elements of the cultural event. Adding other gamification elements (e.g., quizzes, challenges, badges, leaderboards) related to the cultural event may also add value and allow users to enjoy a more direct experience of the metaverse, and effectively convey information about the event. Thus, event managers are encouraged to maximize the use of gamification as an integral part of virtual cultural events; this can engage users and enrich their experiences.

Finally, the results of the analyses underline how important it is that users' can imagine themselves as being in a real cultural event to appreciate its intangible aspects (existential authenticity); this can promote positive behaviors, that is, actual visits to physical events. The metaverse provides users with powerful vicarious "try before you buy" experiences that can enhance their future travel intentions. Affective states have been shown to influence users' ease of imagination and perceptions of authenticity; thus, managers should design specific activities that assist and entertain users during their time attending virtual events, thereby creating a positive mood that will eventually transform metaverse users into visitors to real events.

*6.3. Limitations and future research lines*

This research examined a particular virtual platform that is designed to replicate a specific, real cultural event. This creates limitations that open avenues for future research. First, it has been argued that the current technological status of the platforms is such that none of them meets all the conceptual requirements to be genuinely considered a metaverse (Mystakidis, 2022; Richter & Richter, 2023). Nonetheless, some authors (e.g., Ball, 2022) regard the different proto-metaverses that exist as proxies for the full, future metaverse; this is the standpoint of the authors of the present study. In a similar vein, this research regards the metaverse as a completely generated virtual environment. Other viewpoints see the metaverse as more like a hyperreality, that is, as a fully immersive three-dimensional environment which integrates both physical and virtual worlds through the whole spectrum of immersive technologies (Dwivedi et al., 2022; Koohang et al., 2023). In these fully integrated environments which extend physical reality, AR may play a more important role than VR (Rauschnabel et al., 2022). In other words, the VR-based metaverse does not extend the physical world, it replaces the physical world, whereas the AR-based metaverse extends it through additional information layers and possibilities for interaction. These different perspectives on the technologies that may dominate the metaverse concept underline the need to further explore this field. Therefore, to further analyze the user experience in the metaverse, future studies should take account of how the concept evolves and how the original ideas underlying it are materialized.

Second, to generalize the results future research should examine a wider diversity of metaverse-hosted cultural events. This research examined an event with a pronounced hedonic nature (cultural event). Future studies should analyze whether the effects observed appear in utilitarian events (e.g., conferences, job fairs). As with other immersive technologies, context plays an important role in users' perceptions and behaviors. The congruence between virtual objects and the physical context affects users' experiences (von der Au et al., 2023). This effect is also observed in purely virtual contexts. Thus, different contexts should be taken into account when investigating users' experiences of virtual cultural events held in the metaverse (e.g., hotels, destinations, museums).

This study is among the first to identify and empirically analyze the negative aspects of the user's experience in the metaverse (i.e., lack of focused attention, negative affect), factors that affect the user's ability to envisage, and accept the existential authenticity of, cultural events. It should be noted that other negative aspects of these experiences, such as the lack of real-time multisensory social interactions (Hennig-Thurau et al., 2022), must be considered to understand the current low level of user engagement with the metaverse (Mogaji et al., 2023). In addition, future research should empirically analyze other potential negative consequences of the user experience in the metaverse (e.g., privacy concerns, addiction, diminished reality; Dwivedi et al., 2022, 2023b; Zallio & Clarkson, 2022) and the antecedents and consequences of users' malicious behaviors (e.g., identity theft, harassment; Dwivedi et al., 2023b; Koohang et al., 2023).

Finally, the sampling and design limitations of the studies must be acknowledged. The studies were conducted in Spain, and the samples consisted mainly of young individuals. Future research should increase the geographic and demographic scope of the samples to increase the validity and generalization of the findings. Furthermore, the metaverse was accessed through external devices (computers, smartphones). Future studies should consider whether the use of embodied devices (e.g., VR head-mounted displays; Choi and Kim, 2017) would affect the user experience in the metaverse.

## 7. Conclusions

This study sheds light on the negative effects that the metaverse may have on users' experiences of a virtual cultural event. The many elements integrated into these virtual worlds can distract users from the key elements that make them 'live' the experience. This lack of focused attention generates negative affective states that can impair users' abilities to envisage the cultural event and undermine their belief in its authenticity. Gamification elements can mitigate these negative influences. Gamification activities give users a purpose and make them feel less lost in the metaverse, which improves their experiences. The importance of affective states is evidenced by their direct effects on ease of imagination and perceived authenticity; these effects can turn metaverse users into visitors to real events.

**CRediT authorship contribution statement**

**Carlos Flavián:** Conceptualization, Project administration, Supervision, Writing – Original Draft, Writing – Review & Editing. **Sergio Ibáñez-Sánchez:** Conceptualization, Formal analysis, Methodology, Validation, Writing – Original Draft, Writing – Review & Editing. **Carlos Orús:** Conceptualization, Investigation, Methodology, Supervision, Visualization, Writing – Original Draft, Writing – Review & Editing. **Sergio Barta:** Conceptualization, Formal analysis, Methodology, Validation, Writing – Original Draft, Writing – Review & Editing.

**Declaration of Competing Interest**

After consultation with all authors, Carlos Flavián as corresponding author declares, on behalf of the authors of the paper, that none of the authors has any interest to disclose.

**Acknowledgements**

This study was supported by the Spanish Ministry of Science, Innovation and Universities under Grant PID2019–105468RB-I00 and FPU18/02037; European Social Fund and the Government of Aragon ("METODO" Research Group S20_23R and LMP51_21). We extend our gratitude to the company Imascono Art S.L. for their commitment to this project and for providing the platform utilized in this research.

**Appendix A. Focus group script**

| Instructions | |
|---|---|
| Welcome and focus group explanation | |
| Introduction questions | Name, age, occupation, pre-experience with VR/the metaverse. |
| Examples of event virtualization, metaverse conceptualization | |
| Event virtualizations | 1. Opinion about event virtualization |
| | 2. Event type |
| | 3. Target audience |
| | 4. Reasons for participation |
| Metaverse experience (based on their experience prior to the group session) | 1. Interaction with the environment |
| | 2. Activities conducted |

| Instructions | |
|---|---|
| | 3. Relation to the traditional event |
| | 4. Behavioral intentions after experience |
| New metaverse experience including gamification activities | 1. Differences to experience without gamification |
| | 2. Perceptions |
| | 3. Interaction with the environment |
| | 4. Overall experience |
| Authenticity | 1. Metaverse's capacity to convey authenticity |
| | 2. Impact of gamification |
| Acknowledgments and closing remarks | |

## Appendix B. Scale items

**Lack of focused attention** (α = 0.878; CR=0.910; AVE=0.668)

During my virtual experience…:
… I was fully focused on the online environment (r)
… my attention was focused on surfing in the virtual world (r)
… I was totally concentrated on it (r)
… it was difficult to get information from the site
… it was difficult to concentrate on it

**Negative affect** (α = 0.904; CR=0.934; AVE=0.780)

During my virtual experience, I felt…:
… frustrated
… annoyed
… uncomfortable
… lost

**Ease of imagination** (α = 0.929; CR=0.950; AVE=0.825)

After this virtual experience, it is easier to me to…:
… imagine how the cultural event would be
… imagine myself at the cultural event
… fantasize about being at the cultural event shown
… imagine myself enjoying the cultural event

**Existential authenticity** (α = 0.876; CR=0.910; AVE=0.669)

In this metaverse…:
… I did not feel connected to the history and culture of Zaragoza (r)
… I did not feel the connection with the history and society of Zaragoza (r)
… I felt immersed in the historical atmosphere of the cultural event
… I had a broader knowledge of the cultural tradition of the event
… The cultural event's authenticity is shown

**Behavioral intention** (α = 0.919; CR=0.910; AVE=0.669)

After this virtual experience, I intend to…:
… seek more information about the cultural event
… attend the cultural event
… recommend the cultural event to others

**Perceived gamification** (α = 0.774; CR=0.868; AVE=0.688)

I think this virtual experience…:
… has activities to win prizes
… has competitive challenges and tests
… shows me my achievements

**Knowledge of cultural event** (α = 0.872; CR=0.940; AVE=0.886)

I am informed about the cultural event
I think I know the cultural event well
*I do not need much more information about this cultural event*

**Involvement with cultural event** (α = 0.894; CR=0.934; AVE=0.826)

For me, this cultural event is…:
… Not important – Very important
… Very irrelevant – Very relevant
… Means nothing to me – Means a lot to me

**Place attachment** (α = 0.921; CR=0.945; AVE=0.811)

I am attached to Zaragoza
I feel that Zaragoza is part of me
I feel identified with Zaragoza
I consider myself as a person from Zaragoza

Notes: (r) = reverse items; item in italic was deleted during the validation process.